\input{aipcheck}

\documentclass[
    ,final            % use final for the camera ready runs
  ]
  {aipproc}

\layoutstyle{6x9}

\begin{document}

\title{Evolution of very low mass pre-main sequence stars and young brown dwarfs under accretion\\
\centering{A phenomenological approach}
}
\def\DD{~\mathrm{d}}
\def\LN{\mathrm{\ln}}
\def\msun{M_\odot}
\def\rsun{R_\odot}
\def\mstar{M_\ast}
\def\rstar{R_\ast}
\def\rm{\mathrm}
\def\teff{T_{\mathrm{eff}}}
\def\mdot{\dot{M}}
\def\rv{R_\mathrm{V}}
\def\k{k_\mathrm{B}}
\def\para{=}
\def\inli{\scriptscriptstyle{\rotatebox[origin=c]{90}{\brokenvert}}}
\def\ep{\epsilon}
\def\be{\begin{equation}}
\def\theoretical{{\sf{time dependent accretion rate~}}}
\def\citej#1{\citeauthor{#1}~\citeyear{#1}}
\newcommand{\water}{H$_{2}$O}
\newcommand{\ames}{AMES-H$_2$O}
\newcommand{\phoenix}{\textsc{phoenix}}
\newcommand{\test}{\textit{test-H$_2$O}}
\newcommand{\tio}{AMES-TiO}
\def\ee{\end{equation}}
\def\mjup{$M_\mathrm{{Jup}}$}
\def\ld{L_\mathrm{D}}
\def\spots{{spots}~}
\def\lsun{L_\odot}
\def\xn{\vec{x}^{\,{(n+1)}}}  
\def\xo{\vec{x}^{\,{(n)}}} 
\def\pp#1#2{\frac{\partial#1}{\partial#2}}  


\author{J. Gallardo}{
  address={Departamento de Astronom\'{\i}a, Universidad de Chile, Santiago, Chile}
}

\author{I. Baraffe}{
  address={Ecole Normale Superieure de Lyon, Lyon, France}
}

\author{G. Chabrier}{
  address={Ecole Normale Superieure de Lyon, Lyon, France}
%  ,altaddress={<author1 address>} % additional visiting address
}

\begin{abstract}
In the poster presented in Cool Star 15, we analyzed the effect of disk accretion on the evolution of very low mass pre-main sequence stars and young brown dwarfs and the resulting uncertainties on the determination of masses and ages. We use the Lyon evolutionary 1-D code (\cite{cha}) assuming a magnetospheric accretion process, i.e., the material falls covering a small area of the radiative surface, and we take into account the internal energy added from the accreted material as a free parameter $\epsilon$. Even if the approach to this problem is phenomenological, our formalism provides important hints about characteristics of disk accretion, which are useful for improved stellar interior calculations. Using the accretion rates derived from observations our results show that accretion does not affect considerably the position of theoretical isochrones as well as the luminosity compared with standard non-accreting models (e.g., \cite{ba}).
% In this scenario, we concluded that accretion through a disk cannot reproduce completely the large spread observed in the Hertzsprung-Russell diagram for stellar/sub-stellar objects at an age of 1-3 Myr old. 
See more discussions in a forthcoming paper by Gallardo, Baraffe \& Chabrier (2008).

\end{abstract}

\maketitle

%%%%%%%%%%%%%%%%%%%%%%%%%%%%%%%%%%%%%%%%%%%%
%% MAINMATTER
%%%%%%%%%%%%%%%%%%%%%%%%%%%%%%%%%%%%%%%%%%%%
\vspace{-1cm}
\section{Introduction}

The origin of accretion process is still not well understood, but is it clear that it is strongly linked to the evolution and lifetime of the surrounding circumstellar accreting disk. It appears likely that, in the T-Tauri phase, the inner disk is disrupted by the magnetic field of the central object in which the disk material is channelled along magnetic field lines to crash onto the star (\cite{mu03}). The accreted material falls onto the star in accretion ring or spot, assuming that magnetospheric accretion column covers only a small fraction of the stellar photosphere. 

One of the main physical quantity controlling accretion material through the disk is the mass accretion rate, $\mdot$. Recent observations of young low mass stars and brown dwarfs formation regions derived accretion rates increasing with the stellar masses: $\mdot \propto \mstar^2$ (\cite{natta}; \cite{mo}). 
On the other hand, the derived sub and/or stellar masses depend on evolutionary tracks. However, these evolutionary models at young ages suffer from several uncertainties, mainly due to unknown initial conditions resulting from prior cloud collapse and accretion phases. The evolutionary tracks need to be modified to take into account accretion process that build up stars, 

The purpose of this poster is to present theoretical evolutionary models analyzing the effect of accretion on the evolution of young very low mass pre-main sequence stars and young brown dwarfs and the resulting uncertainties on the determination of masses and ages, based on detailed input physics describing the structure and evolution of such low mass objects and to extend it to the brown dwarfs regime. One of the main idea is to analyze the sensitivity of evolutionary tracks to different accretion rates in the stellar sub-stellar mass regime. 
\vspace{-0.8cm}
\section{Results}
%%%%%%%%%%%%%%%%%%%%%%%%%%%%%%%%%%%%%%%%%%%%
%% Sample figure:
%%
%% The option [height=...] scales the picture to the given height,
%% without it it would be printed at its nominal size
%%%%%%%%%%%%%%%%%%%%%%%%%%%%%%%%%%%%%%%%%%%%

%\begin{figure}
%  \includegraphics[angle=270,height=.3\textheight]{fig1.eps}
%  \caption{}
%\end{figure}

\begin{figure}[h!]
\centering
\includegraphics
%[height=8cm,
[width=6.cm,
angle=-90]
{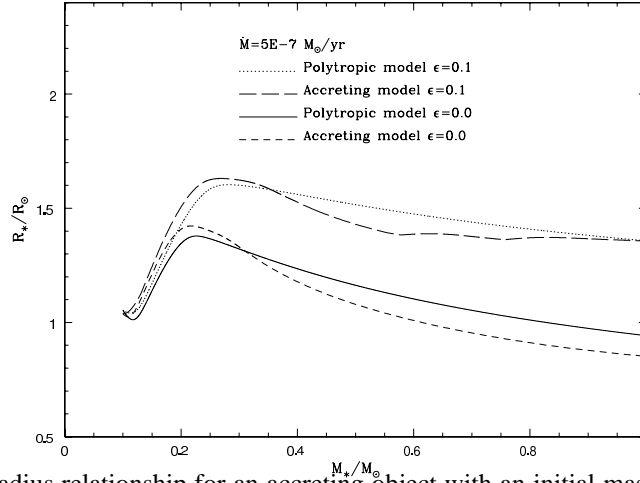}
\caption{Mass-radius relationship for an accreting object with an initial mass of 0.1 solar masses, constant accretion rate of $dM/dt$=5E-7 $M_\odot$ and different $\epsilon$ values (showed in the inset). The polytropic model  corresponds to the \cite{hck} formalism. The internal energy $U$ is added to the accreting object as $\Delta U \propto \epsilon \Delta M$, where M  is the object's mass.}
\label{}
\end{figure}

%#############################

\begin{figure}[h!]
\centering
\includegraphics
%[height=8cm,
[width=8cm,
angle=0]
{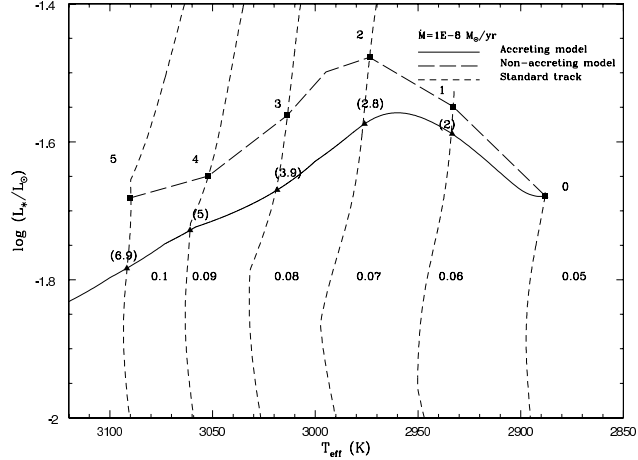}
\caption{Hertzsprung-Russell diagram (HRD) for an accreting object with initial mass of 0.05 $M_\odot$ and $\epsilon=0.0$ (i.e., without internal energy added to the object) represented in solid line. The vertical short dashed lines are cooling tracks for non-accreting low-mass objects, with masses indicated near the curves (in $M_\odot$). The full squares indicate the position of non-accreting objects with the same age (indicated by the numbers close to the squares, in Myr) and same mass than  accreting counterpart (indicated by the triangles just below the squares). The numbers in brackets (close to the triangles) give the age, in Myr, of non-accreting objects at the position indicated by the triangles. We see that assigning an age for an observed young object with a given luminosity from non accreting tracks, we can significantly overestimate its age. This illustrates the uncertainty in age determination based on standard evolutionary tracks at young ages.
}
\label{}
\end{figure}

%############################

\begin{figure}[h!]
\centering
\includegraphics
%[height=8cm,
[width=9.0cm,
angle=0]
{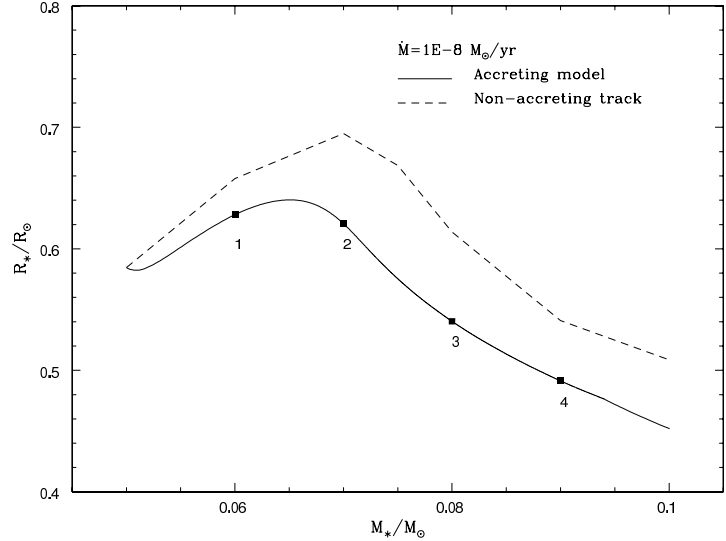}
\caption{Effect of accretion on the mass-radius relationship. Solid line shows an accreting object with initial mass of 0.05 $M_\odot$. The accreting model was calculated for $\epsilon=0.0$. The short dashed line indicates the radius of non-accreting objects with the same mass and the same age as their accreting counterparts. Ages for the accreting object, in Myr, are indicated by the numbers close to the squares. At any time the structure of the accreting object is more compact than its non-accreting counterpart, i.e, accretion time-scale becomes of the order of Kelvin-Helmholtz time-scale, thus the accreting object does not have time to expand to the radius at the corresponding mass it would have in the absence of accretion.
}
\label{}
\end{figure}

\begin{figure}[h!]
\centering
\includegraphics
%[height=8cm,
[width=6.5cm,
angle=-90]
{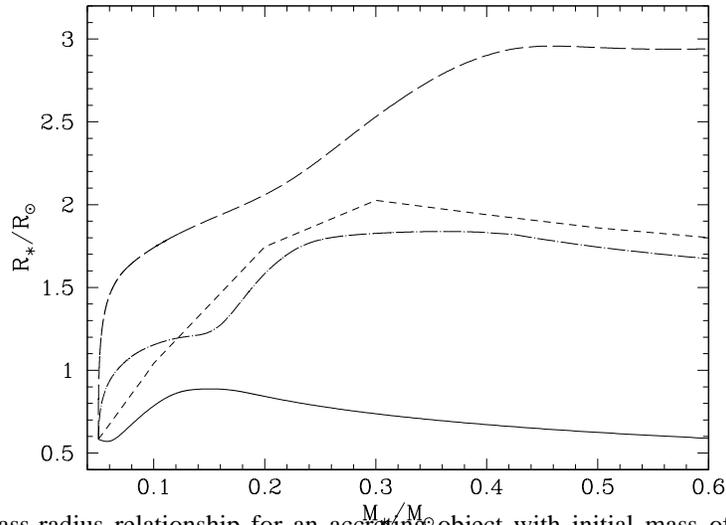}
\caption{Mass-radius relationship for an accreting object with initial mass of 0.05 $M_\odot$ and an accretion rate of 5e-7 $M_\odot/yr$ (for all showed models). The solid line displays a model using $\epsilon=0.0$, dotted long dashed line shows a model for $\epsilon=0.1$, and long dashed line portrays a model with $\epsilon=0.5$. The short dashed line correspond to the short dashed line in Fig.~3. Following the discussion by \cite{sta}, the accreting object expand in radius and it evolves maintaining a nearly fixed mass-radius ratio as long as deuterium (D) burning nuclear energy is sufficiently high. The effect of $\epsilon \neq 0$ makes the star grow in radius until the extra supply of D becomes ineffective, thus the luminosity loss is dominant and the object need to contract.
}
\label{}
\end{figure}

\begin{figure}[h!]
\centering
\includegraphics
%[height=8cm,
[width=6.8cm,
angle=0]
{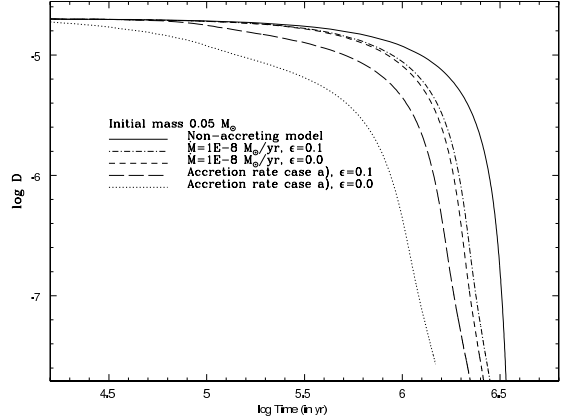}
\caption{Evolution of deuterium abundance for an accreting object of 0.05 $M_\odot$. Solid line shows a non-accreting model. The dotted long dashed and short dashed lines are models calculated using $dM/dt$=1E-8 $M_\odot/yr$ and $\epsilon=0.0$ and $\epsilon=0.1$ respectively. The long dashed and dotted lines are models using $dM/dt$=1E-6 $M_\odot/yr$ and $\epsilon=0.0$ and $\epsilon=0.1$ respectively. The additional internal energy compensates the energy loss by radiation at the surface, because the internal energy input reduces the need for deuterium burning to supply the entire luminosity.
}
\label{}
\end{figure}

\begin{figure}[h!]
\centering
\includegraphics
%[height=8cm,
[width=6.8cm,
angle=0]
{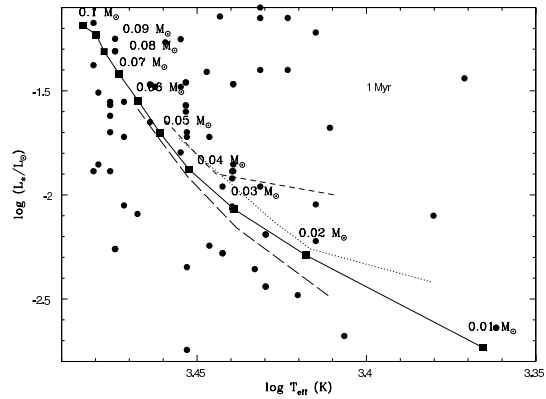}
\caption{HRD for young low mass objects and several theoretical isochrones taking into account different accretion rates and $\epsilon$ values. Data were taken mostly from \cite{mu05}, \cite{lu} and \cite{natta} showed in filled circles. Standard non-accreting isochrone from 0.01 to 0.1 $M_\odot$ (indicated by the numbers close to the squares) is displayed in solid line. The long dashed line shows a model using $dM/dt$=1E-8 $M_\odot/yr$ and $\epsilon=0.0$. Dotted line shows a model using $dM/dt$=5E-7 $M_\odot/yr$ and $\epsilon=0.3$. Short dashed line corresponds to a model using $dM/dt$=1E-7 $M_\odot/yr$ and $\epsilon=0.5$. All models were calculated for 1 Myr old. Data age range from 1 to 3 Myr old. 
}
\label{}
\end{figure}

%\endinput
\end{document}